%% file: ICRC2023_template_IceCube.tex
\documentclass[a4paper,11pt]{article}

\usepackage{lineno}

\usepackage{pos}
\usepackage{lipsum} 

\title{Search for neutrino sources from the direction of IceCube alert events 
}

\ShortTitle{Search for neutrino sources from the direction of IceCube alert events}

\author{The IceCube Collaboration \\{\normalsize \normalfont(a complete list of authors can be found at the end of the proceedings)}\\}

\emailAdd{martina.karl@icecube.wisc.edu}

\abstract{

We search for additional neutrino emission from the direction of IceCube's highest energy public alert events. We take the arrival direction of 122 events with a high probability of being of astrophysical origin and look for steady and transient emission. We investigate 11 years of reprocessed and recalibrated archival IceCube data. For the steady scenario, we investigate if the potential emission is dominated by a single strong source or by many weaker sources. In contrast, for the transient emission we only search for single sources. In both cases, we find no significant additional neutrino component. Not having observed any significant excess, we constrain the maximal neutrino flux coming from all 122 origin directions (including the high-energy events) to $\Phi_{90\%,~100~\rm{TeV}} = 1.2 \times 10^{-15}$~(TeV cm$^2$ s)$^{-1}$ at 100~TeV, assuming an $E^{-2}$ emission, with 90\% confidence.
The most significant transient emission of all 122 investigated regions of interest is the neutrino flare associated with the blazar TXS~0506+056. With the recalibrated data, the flare properties of this work agree with previous results. We fit a Gaussian time profile centered at $\mu_T = 57001 ^{+38}_{-26}$~MJD and with a width of $\sigma_T = 64 ^{+35}_{-10}$~days. The best fit spectral index is $\gamma = 2.3 \pm 0.4$ and we fit a single flavor fluence of $J_{100~\rm{TeV}} = 1.2 ^{+1.1} _{-0.8} \times 10^{-8} $~(TeV~cm$^2$)$^{-1}$. The global p-value for transient emission is $p_{\rm{global}} = 0.156$ and, therefore, compatible with background.

\vspace{4mm}
{\bfseries Corresponding authors:}
Martina Karl$^{*1,2} $\\
{$^{1}$ \itshape Technical University of Munich, TUM School of Natural Sciences, Department of Physics, James-Franck-Straße 1, D-85748 Garching bei M\"unchen, Germany}\\
{$^{2}$ \itshape European Southern Observatory, Karl-Schwarzschild-Straße 2, D-85748 Garching bei M\"unchen, Germany}\\[4mm]
$^*$ Presenter

\ConferenceLogo{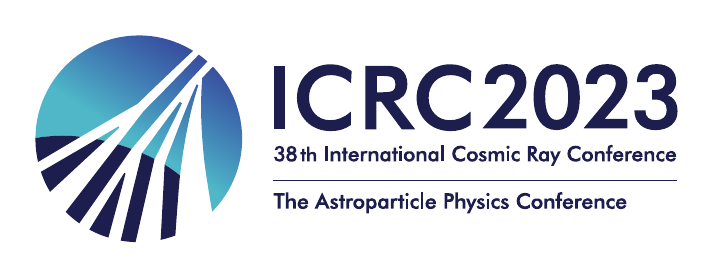}

\FullConference{The 38th International Cosmic Ray Conference (ICRC2023)\\ 26 July -- 3 August, 2023\\ Nagoya, Japan}
}

\begin{document}

\maketitle

\section{Introduction}\label{sec:intro}

The IceCube Neutrino Observatory \cite{Aartsen_2017} instruments a cubic kilometer of Antarctic ice at the geographic South Pole. It has a high duty cycle ($> 99\%$) \cite{Aartsen_2017}, and contrary to traditional astronomical telescopes, IceCube's field of view covers the full sky while being most sensitive to high-energy events at its horizon. Hence, IceCube is ideal for monitoring the sky for interesting events and alerting other telescopes. If IceCube detects neutrino events with a high probability of being of astrophysical origin (i.e., high energies and good pointing), it alerts other telescopes \cite{2017APh....92...30A} and triggers follow-up multi-messenger observations \cite{Kintscher_2016}. On the 22nd of September 2017, such a high-energy event was detected (IceCube-170922A), and the astronomical community observed a flaring blazar at the reconstructed origin direction of IceCube-170922A \cite{2018Sci...361.1378I}. Triggered by this association, a follow-up search for time-dependent neutrino emission before IceCube-170922A found a neutrino flare probably originating from TXS~0506+056 between September 2014 and March 2015 \cite{2018Sci...361..147I}. 

\begin{figure}[h]
    \centering
    \includegraphics[width=0.7\textwidth]{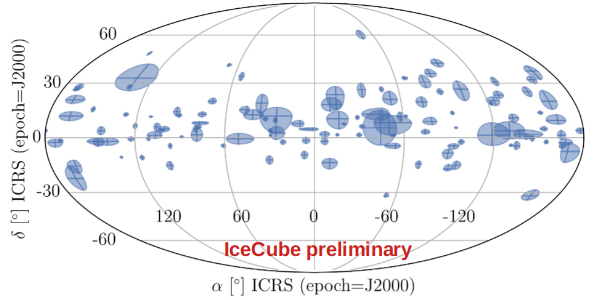}
    \caption{Map of all high-energy tracks and their 90\% uncertainty regions investigated in this work. Events were detected between 2009 and the end of 2021.}
    \label{fig:skymap}
\end{figure}

We investigate if, in general (so without an electromagnetic counterpart), there is an association between high-energy alerts and additional neutrino emission, as was the case for IceCube-170922A and the neutrino flare. Preliminary results were presented in \cite{2019ICRC...36..929K}; however, the selection for high-energy events was updated in the meantime \cite{2022ApJ...928...50A}. We choose alert events with the highest astrophysical purity (the ``gold'' alert channel \cite{blaufuss2019generation, abbasi2023icecat1}), add high-energy events from Ref. \cite{2022ApJ...928...50A}, and use their origin directions as positions of interest for a neutrino follow-up analysis. Since these alert events motivate the analysis, we remove the respective events from the analyzed data. Hence, we investigate whether these highest-energy events are generally associated with neutrino emission in lower energies. Figure \ref{fig:skymap} shows all the analyzed positions in this work and their uncertainty regions. For the analysis, we use 11 years of re-calibrated (with respect to Ref. \cite{2018Sci...361..147I}) through-going muon tracks, from mid 2009 to mid 2020. 

\section{Analysis method}\label{sec:method}

We use the unbinned likelihood ratio approach presented in Ref. \cite{Braun:2008bg}. The likelihood, $\mathcal{L}$, is the joint probability of the signal, $S$, and the background, $B$, probability density function (pdf). The likelihood ratio test then evaluates the ratio of a signal and a background hypothesis. The background hypothesis assumes no emission from neutrino point sources; hence the data are atmospheric background and a diffuse astrophysical neutrino component. The signal hypothesis states that there is a point-source signal of $n_S$ neutrinos on top of the expected background. The neutrinos cluster around the source location and follow a power-law spectrum $\propto E^{-\gamma}$. In the case of a time-dependent emission, we expect neutrinos only during a certain period of time (assuming a Gaussian temporal pdf). 

With the likelihood ratio, we express the test statistic value as
\begin{equation}\label{eq:TS_timeint_exp}
\text{TS} = 2 \ln  \left[\frac{\mathcal{L}(n_S=\hat{n}_S)}{\mathcal{L}(n_S=0)} \right] 
 = 2 ~ \sum_{i} \ln \left[\frac{\hat{n}_S}{N}\left(\frac{S_i(\hat{\gamma})}{B_i}-1\right) +1\right],
\end{equation}
where we optimize the mean number of detected signal neutrinos, $n_S$, and the source spectral index, $\gamma$ (as part of the signal pdf $S$). The sum over background and signal pdfs is taken over all events, $i=1--N$. The $\hat{n}_S, \hat{\gamma}$ indicate the best-fit values of $n_S$ and $\gamma$.

This analysis searches for continuous and for time-dependent emission from the direction of IceCube alert events. For the continuous case, the signal and background pdfs are the product of a spatial and energy pdf. For the signal pdf, the spatial part depends on the distance between the source position, $\Vec{x}_S$, and reconstruction position of the neutrino events, $\vec{x}_i$ and their reconstruction uncertainty, $\sigma_i$. The signal energy pdf describes the probability of detecting an event with energy $E_i$ at declination $\delta_i$ originating from a source with energy spectrum $\propto E^{-\gamma}$

\begin{equation}\label{eq:signal_timeint_gen}
S_i (\vec{x}_i ,  E_i | \sigma_i, \vec{x}_S, \gamma) = S_{\text{spatial}} (\vec{x}_i| \sigma_i , \vec{x}_S) \cdot S_{\text{energy}} (E_i | \delta _i , \gamma ) 
= \frac{1}{2 \pi \sigma_i ^2} \exp \left( \frac{- | \vec{x}_i - \vec{x}_S | ^2}{2 \sigma _i ^2} \right) \cdot S_{\text{energy}} (E_i | \delta _i, \gamma ) .
\end{equation}

For the background, we can assume uniformity in right ascension due to IceCube's unique position at the South Pole

\begin{equation}\label{eq:bg_timeint_exp}
B_i \left( \vec{x}_i, E_i \right) = B_{\text{spatial}} (\vec{x}_i) \cdot B_{\text{energy}} (E_i| \delta _i) = \frac{1}{2 \pi} \cdot P(\delta _i) \cdot B_{\text{energy}}(E_i| \delta _i).
\end{equation}

In the case of the time-dependent analysis, the signal and background pdfs are the product of a spatial, energy, and temporal pdf. The spatial and energy pdfs remain the same as for the continuous case. For the signal temporal pdf, $S_T$, we assume a Gaussian-shaped time profile 

\begin{equation}\label{eq:sig_pdf_temporal_gauss}
S_{T}(t_i | \mu_T, \sigma_T) = \frac{1}{{\sigma_T \sqrt {2\pi } }}\exp \left( \frac{- \left( {t_i - \mu_T } \right)^2 } {2\sigma_T ^2 } \right),
\end{equation}

centered at $\mu_T$ with width $\sigma_T$ and evaluate the pdf for the detection time of each event, $t_i$. For the background, the temporal pdf assumes uniformity over the whole detection time: $1/\text{detector uptime}$. When investigating possible flaring times, short flares are more abundant than longer flares. This introduces a bias towards shorter flares, which we correct by multiplying the test statistic with $\sigma_T / 300~\text{days}$, where 300 days is the upper bound on the flare length. 

However, we do not know when the source could be flaring. We used, for the first time on IceCube data, the unsupervised learning algorithm expectation maximization (EM) \cite{10.2307/2984875} to identify neutrino flares \cite{karl2021search, dissertation_martina}. EM is based on a two-component Gaussian mixture model; hence we assume a Gaussian signal and a uniform background. We weigh each event time with their time-independent signal pdf divided by the background pdf ($S_i(\gamma) / B_i)$ as in equation \ref{eq:signal_timeint_gen} and equation \ref{eq:bg_timeint_exp}. Since $S_i(\gamma)$ depends on the source spectral index, the weights ($S_i(\gamma) / B_i)$ change depending which $\gamma$ we assume for a source. To avoid biases for a specific spectral index, we calculate the weights for different spectral indices and repeat the fit of the neutrino flare with EM for each. For each best fit $\hat{\mu}_T$ and $\hat{\sigma}_T$, we calculate the TS value (this time including the temporal pdfs) and take the parameters yielding the best (i.e., highest) TS value as the best-fit parameters. 

The above procedure assumes a precise position $\vec{x}_S$. However, as shown in Figure \ref{fig:skymap}, there are uncertainties on the alert events' reconstructed positions. Hence, we fit the best point-source position within each uncertainty region. We divide each region into a grid with steps of $0.2^{\circ}$ and calculate each grid point's TS-value (by optimizing $n_S$ and $\gamma$). $0.2^{\circ}$ is the best possible angular resolution and smaller than the median angular resolution. The point yielding the most significant TS-value is then the best-fit point-source position. We then compare this TS-value with a background TS distribution generated for each region and calculate the local p-value (i.e., the probability of getting this TS-value if the data were background-like). 

We do this analysis for 122 uncertainty regions and choose the most significant p-value as our result. However, we have to correct for the fact we are investigating 122 regions. We run the analysis $\sim 10^4$ times on background data, and each time we take the most significant p-value out of the 122 analyzed regions. By comparing these background p-values with the one we obtain from the real data, we get the global p-value (i.e., the probability of getting the most significant local p-value if the data were background-like). 

In the continuous case, we additionally investigate if there is an overall neutrino emission from all 122 regions combined. For this, we adopt the approach by Ref. \cite{Albert_2022} and sum the TS-values of all 122 regions and get a stacked TS-value. The same is done for $~10^4$ background realizations, and we calculate a stacked p-value for overall emission from all positions based on the background distribution. The stacked p-values are already the global p-values and need no further correction.

\section{Results}\label{sec:results}

Our search for continuous neutrino emission from the direction of IceCube alert events (while not including the respective alert event in the data) yields a global p-value of 98\%. Hence, we find no evidence for an additional continuous component to the single high-energy events. The search for the overall emission from all regions combined yields a p-value of 8\%, which is still compatible with the background expectation. Hence, we constrain the possible overall neutrino emission from all regions combined (excluding the alert events) to $\Phi_{90\%, 100\rm{TeV}}^{\nu_{\mu} + \bar{\nu}_{\mu}, \rm{w/o~alerts}} = 4.2 \times 10^{-16}~(\rm{TeV}~\rm{cm}^2~\rm{s})^{-1}$ between 4.2~TeV and 3.6~PeV with 90\% confidence. Here, we assume a spectral emission of $\propto E^{-2}$ and the flux is normalized at 100~TeV.
For comparison with the diffuse flux, we include the alert events in the analysis and determine a 90\% confidence upper flux limit of $\Phi_{90\%,100\rm{TeV}}^{\nu_{\mu} + \bar{\nu}_{\mu}, \rm{with~alerts}}= 1.2 \times 10^{-15}~(\rm{TeV}~\rm{cm}^2~\rm{s})^{-1}$ for a total emission of these high-energy neutrino production sites. Here again, we assume an emission $\propto E^{-2}$. Comparing this with the astrophysical diffuse flux of $\Phi_{\text{diffuse},100\rm{TeV}} = 1.44 \times 10^{-15}~(\rm{TeV}~\rm{cm}^2~\rm{s}~\rm{sr})^{-1} $ between 15~TeV and 5~PeV \citep{2022ApJ...928...50A}, we conclude that the emission from all these regions can contribute a maximum of 4.6 \% of the astrophysical diffuse flux (between 15~TeV and 3.6~PeV). A more detailed discussion on the continuous results can be found in \cite{dissertation_martina}.

\begin{figure}
    \centering
    \includegraphics[width=0.5\textwidth]{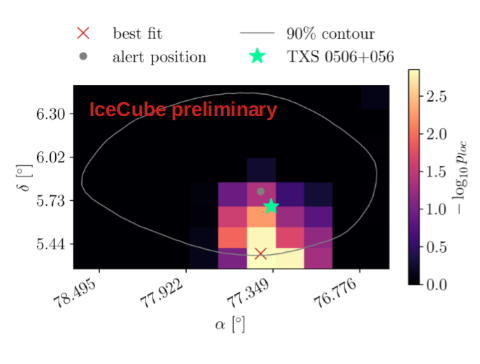}
    \caption{P-value map of the uncertainty region of IceCube-170922A. The grey dot shows the best-reconstructed direction of IceCube-170922A, the grey line the 90\% uncertainty contour on the reconstruction. The red cross shows the best-fit position, about $0.5^{\circ}$ of TXS~0506+056 (green star). The color displays the local p-value at each gridpoint.}
    \label{fig:txs_skyscan}
\end{figure}

\begin{figure}
    \centering
    \includegraphics[width=0.9\textwidth]{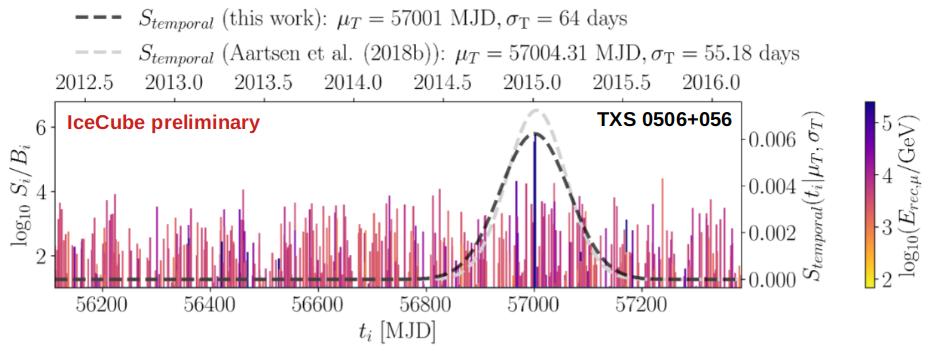}
    \caption{Events' $S_i / B_i$ versus detection time, $t_i$ for TXS~0506+056. The black line shows this work's temporal pdf, and it agrees with the grey line showing the temporal pdf published in Ref. \cite{2018Sci...361..147I}. The event color shows their reconstructed muon energy.}
    \label{fig:txs_sob}
\end{figure}

\begin{figure}
    \centering
    \includegraphics[width=0.8\textwidth]{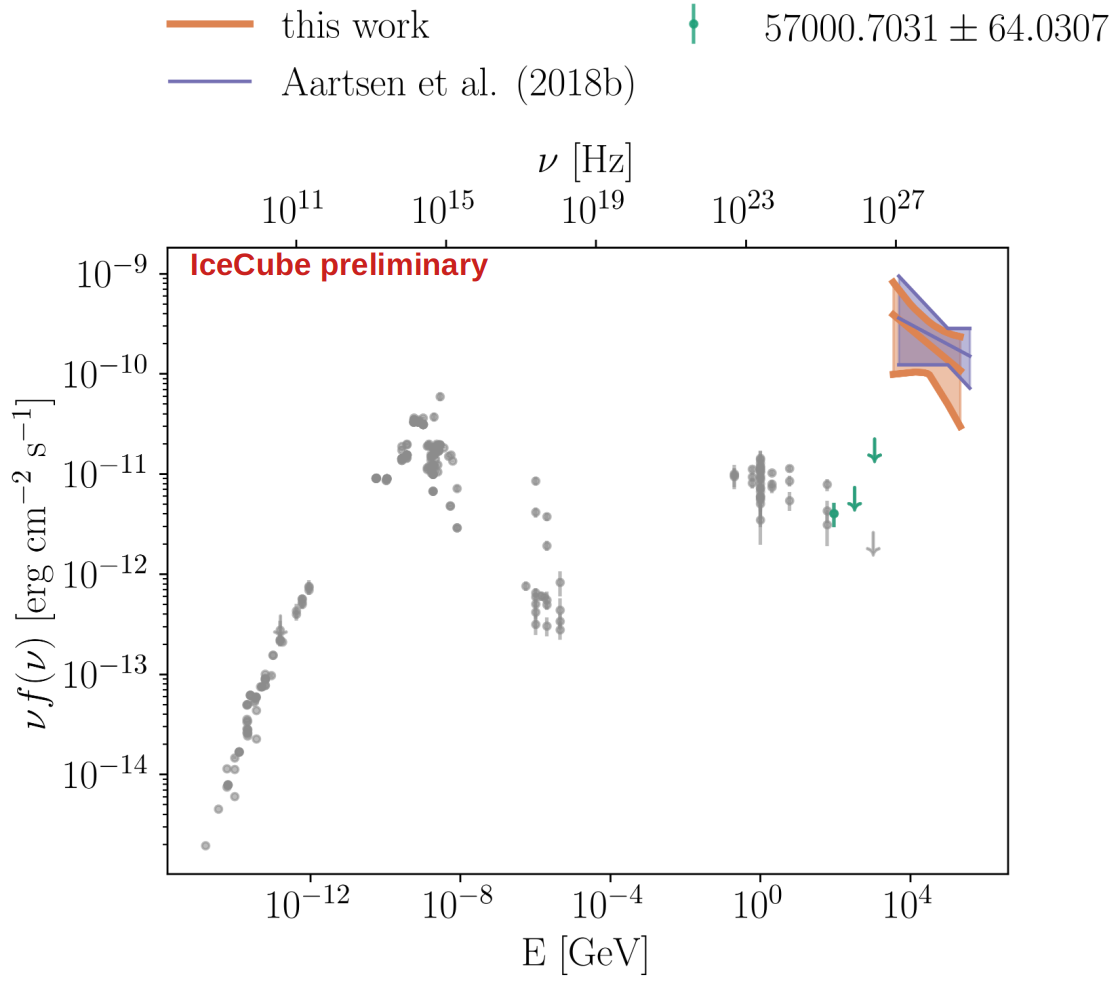}
    \caption{Spectral energy distribution of TXS~0506+056. Photons are shown in grey dots, the green dots in $\gamma$-rays were emitted during the time of the neutrino flare ($57001 \pm 64$~days). The neutrino band compares this work (orange) with the result of Ref. \cite{2018Sci...361..147I} (purple). }
    \label{fig:txs_sed}
\end{figure}

When looking for time-dependent sources, the most significant region is from the alert IceCube-170922A, which is associated with the blazar TXS~0506+056. The best-fit position is within $0.5^{\circ}$ of the location of TXS~0506+056 (see Figure \ref{fig:txs_skyscan}). The local p-value is $3\,\sigma$ and the best-fit parameters agree with Ref. \cite{2018Sci...361..147I}, as shown in Figure \ref{fig:txs_sob}. We find a flare with mean $\hat{\mu}_T = 57001_{-26}^{+38}$~MJD and a width of $\hat{\sigma}_T=64_{-10}^{+35}$~days (agreeing with $\hat{\mu}_T = 57004_\pm 21$~MJD and a width of $\hat{\sigma}_T=55_{-12}^{+18}$~days in Ref. \cite{2018Sci...361..147I}). The best-fit mean number of expected neutrinos during the flare is $\hat{n}_S = 12^{+9}_{-7}$, with an energy spectral index of $\hat{\gamma} = 2.3 \pm 0.4$. During the flare ($\hat{\mu}_T \pm 2\hat{\sigma}_T$), the average flux is $\Phi_{100\rm{TeV}}^{\nu_{\mu} + \bar{\nu}_{\mu}} = 1.1 ^{+0.9}_{-0.8} \times 10^{-15}~(\rm{TeV}~\rm{cm}^2~\rm{s})^{-1}$ between 3.5~TeV and 213~TeV. The time-integrated flux (integrated over $\hat{\mu}_T \pm 2\hat{\sigma}_T$), the fluence, is $J_{100\rm{TeV}}^{\nu_{\mu} + \bar{\nu}_{\mu}} = 1.2 ^{+1.1} _{-0.8} \times 10^{-8} $~(TeV~cm$^2$)$^{-1}$. The flux estimates and spectral index agree with Ref. \cite{2018Sci...361..147I} ($\Phi_{100\rm{TeV}}^{\nu_{\mu} + \bar{\nu}_{\mu}} = 1.6 ^{+0.7}_{-0.6} \times 10^{-15}~(\rm{TeV}~\rm{cm}^2~\rm{s})^{-1}$ and $\hat{\gamma} = 2.2 \pm 0.2$), as shown in Figure \ref{fig:txs_sed}. Details on the time-dependent analysis, how the re-calibration affects the flare, and the uncertainty estimations can be found in \cite{dissertation_martina}. 

\section{Conclusion}\label{sec:conclusion}
In general, high-energy alert events do not point to lower-energy neutrino emission. The lack of lower-energetic neutrino emission (with respect to IceCube alert events) could be, for example the result of sources emitting a very hard energy spectrum $\left(\propto E^{0 < \gamma \leq -1}\right)$. We would see such a hard spectrum as single high-energy events since lower energetic events would be sparse and completely dominated by atmospheric background. Another possibility are flaring sources where we only detect single high-energy events during the flare. Considering TXS~0506+056, it could be that there are similar sources that were not flaring in lower energies between mid-2009 and mid-2020. All in all, TXS~0506+056 remains the only locally significant source, and the alert event IceCube-170922A is quite unique in its association with lower-energy neutrinos.

\bibliographystyle{ICRC}
\bibliography{references}

%

\clearpage

\input{authorlist_IceCube.tex}

\end{document}

%% file: authorlist_IceCube.tex
\section*{Full Author List: IceCube Collaboration}

\scriptsize
\noindent
R. Abbasi$^{17}$,
M. Ackermann$^{63}$,
J. Adams$^{18}$,
S. K. Agarwalla$^{40,\: 64}$,
J. A. Aguilar$^{12}$,
M. Ahlers$^{22}$,
J.M. Alameddine$^{23}$,
N. M. Amin$^{44}$,
K. Andeen$^{42}$,
G. Anton$^{26}$
% ...
\\
\\
% ...
$^{12}$ Universit{\'e} Libre de Bruxelles, Science Faculty CP230, B-1050 Brussels, Belgium \\
% ...
$^{17}$ Department of Physics, Loyola University Chicago, Chicago, IL 60660, USA \\
$^{18}$ Dept. of Physics and Astronomy, University of Canterbury, Private Bag 4800, Christchurch, New Zealand \\
% ...
$^{22}$ Niels Bohr Institute, University of Copenhagen, DK-2100 Copenhagen, Denmark \\
$^{23}$ Dept. of Physics, TU Dortmund University, D-44221 Dortmund, Germany \\
% ...
$^{26}$ Erlangen Centre for Astroparticle Physics, Friedrich-Alexander-Universit{\"a}t Erlangen-N{\"u}rnberg, D-91058 Erlangen, Germany \\
% ...
$^{40}$ Dept. of Physics and Wisconsin IceCube Particle Astrophysics Center, University of Wisconsin{\textendash}Madison, Madison, WI 53706, USA \\
% ...
$^{42}$ Department of Physics, Marquette University, Milwaukee, WI, 53201, USA \\
% ...
$^{44}$ Bartol Research Institute and Dept. of Physics and Astronomy, University of Delaware, Newark, DE 19716, USA \\
% ...
$^{63}$ Deutsches Elektronen-Synchrotron DESY, Platanenallee 6, 15738 Zeuthen, Germany  \\
$^{64}$ Institute of Physics, Sachivalaya Marg, Sainik School Post, Bhubaneswar 751005, India \\
% ...

\subsection*{Acknowledgements}

\noindent
The authors gratefully acknowledge the support from the following agencies and institutions:
USA {\textendash} U.S. National Science Foundation-Office of Polar Programs,
U.S. National Science Foundation-Physics Division,
U.S. National Science Foundation-EPSCoR,
Wisconsin Alumni Research Foundation,
Center for High Throughput Computing (CHTC) at the University of Wisconsin{\textendash}Madison,
Open Science Grid (OSG),
Advanced Cyberinfrastructure Coordination Ecosystem: Services {\&} Support (ACCESS),
Frontera computing project at the Texas Advanced Computing Center,
U.S. Department of Energy-National Energy Research Scientific Computing Center,
Particle astrophysics research computing center at the University of Maryland,
Institute for Cyber-Enabled Research at Michigan State University,
and Astroparticle physics computational facility at Marquette University;
Belgium {\textendash} Funds for Scientific Research (FRS-FNRS and FWO),
FWO Odysseus and Big Science programmes,
and Belgian Federal Science Policy Office (Belspo);
Germany {\textendash} Bundesministerium f{\"u}r Bildung und Forschung (BMBF),
Deutsche Forschungsgemeinschaft (DFG),
Helmholtz Alliance for Astroparticle Physics (HAP),
Initiative and Networking Fund of the Helmholtz Association,
Deutsches Elektronen Synchrotron (DESY),
and High Performance Computing cluster of the RWTH Aachen;
Sweden {\textendash} Swedish Research Council,
Swedish Polar Research Secretariat,
Swedish National Infrastructure for Computing (SNIC),
and Knut and Alice Wallenberg Foundation;
European Union {\textendash} EGI Advanced Computing for research;
Australia {\textendash} Australian Research Council;
Canada {\textendash} Natural Sciences and Engineering Research Council of Canada,
Calcul Qu{\'e}bec, Compute Ontario, Canada Foundation for Innovation, WestGrid, and Compute Canada;
Denmark {\textendash} Villum Fonden, Carlsberg Foundation, and European Commission;
New Zealand {\textendash} Marsden Fund;
Japan {\textendash} Japan Society for Promotion of Science (JSPS)
and Institute for Global Prominent Research (IGPR) of Chiba University;
Korea {\textendash} National Research Foundation of Korea (NRF);
Switzerland {\textendash} Swiss National Science Foundation (SNSF);
United Kingdom {\textendash} Department of Physics, University of Oxford.

%% file: ICRC2023_template_IceCube.bbl
\providecommand{\href}[2]{#2}\begingroup\raggedright\begin{thebibliography}{10}

\bibitem{Aartsen_2017}
{\bfseries IceCube} Collaboration, M.~Aartsen {\em et~al.}
  \href{http://dx.doi.org/10.1088/1748-0221/12/03/p03012}{{\em Journal of
  Instrumentation} {\bfseries 12} no.~03, (2017) P03012--P03012}.

\bibitem{2017APh....92...30A}
{\bfseries IceCube} Collaboration, M.~G. Aartsen {\em et~al.}
  \href{http://dx.doi.org/10.1016/j.astropartphys.2017.05.002}{{\em Astropart.
  Phys.} {\bfseries 92} (2017) 30--41}.

\bibitem{Kintscher_2016}
{Thomas Kintscher for the IceCube Collaboration}
  \href{http://dx.doi.org/10.1088/1742-6596/718/6/062029}{{\em Journal of
  Physics: Conference Series} {\bfseries 718} (2016) 062029}.

\bibitem{2018Sci...361.1378I}
{\bfseries IceCube, Fermi-LAT, MAGIC, AGILE, ASAS-SN, HAWC, H.E.S.S, INTEGRAL,
  Kanata, Kiso, Kapteyn, Liverpool telescope, Subaru, Swift/NuSTAR, VERITAS,
  and VLA/17b-403} Collaboration, M.~G. {Aartsen} {\em et~al.}
  \href{http://dx.doi.org/10.1126/science.aat1378}{{\em Science} {\bfseries
  361} (2018) eaat1378}.

\bibitem{2018Sci...361..147I}
{\bfseries IceCube} Collaboration, M.~G. {Aartsen} {\em et~al.}
  \href{http://dx.doi.org/10.1126/science.aat2890}{{\em Science} {\bfseries
  361} (2018) 147--151}.

\bibitem{2019ICRC...36..929K}
M.~{Karl}, ``{Search for Neutrino Emission in IceCube's Archival Data from the
  Direction of IceCube Alert Events},'' in {\em 36th International Cosmic Ray
  Conference (ICRC2019)}, vol.~36 of {\em International Cosmic Ray Conference},
  p.~929.
\newblock July, 2019.
\newblock \href{http://arxiv.org/abs/1908.05162}{{\ttfamily arXiv:1908.05162
  [astro-ph.HE]}}.

\bibitem{2022ApJ...928...50A}
{\bfseries IceCube} Collaboration, R.~{Abbasi}, M.~{Ackermann}, J.~{Adams},
  J.~A. {Aguilar}, {\em et~al.}
  \href{http://dx.doi.org/10.3847/1538-4357/ac4d29}{{\em Astrophysical Journal}
  {\bfseries 928} no.~1, (Mar., 2022) 50}.

\bibitem{blaufuss2019generation}
E.~{Blaufuss}, T.~{Kintscher}, L.~{Lu}, and C.~F. {Tung},
  \href{http://dx.doi.org/10.22323/1.358.01021}{``{The Next Generation of
  IceCube Real-time Neutrino Alerts},''} in {\em 36th International Cosmic Ray
  Conference (ICRC2019)}, vol.~36 of {\em International Cosmic Ray Conference},
  p.~1021.
\newblock July, 2019.
\newblock \href{http://arxiv.org/abs/1908.04884}{{\ttfamily arXiv:1908.04884
  [astro-ph.HE]}}.

\bibitem{abbasi2023icecat1}
{\bfseries IceCube} Collaboration, R.~Abbasi {\em et~al.}, ``Icecat-1: the
  icecube event catalog of alert tracks,'' 2023.

\bibitem{Braun:2008bg}
J.~Braun, J.~Dumm, F.~De~Palma, C.~Finley, A.~Karle, and T.~Montaruli
  \href{http://dx.doi.org/10.1016/j.astropartphys.2008.02.007}{{\em Astropart.
  Phys.} {\bfseries 29} (2008) 299--305}.

\bibitem{10.2307/2984875}
A.~P. Dempster, N.~M. Laird, and D.~B. Rubin {\em Journal of the Royal
  Statistical Society. Series B (Methodological)} {\bfseries 39} no.~1, (1977)
  1--38.

\bibitem{karl2021search}
M.~Karl, P.~Eller, and A.~Schubert,
  \href{http://dx.doi.org/10.22323/1.395.0940}{``{Search for high-energy
  neutrino sources from the direction of IceCube alert events},''} in {\em
  Proceedings of 37th International Cosmic Ray Conference {\textemdash}
  PoS(ICRC2021)}, vol.~395, p.~940.
\newblock 2021.

\bibitem{dissertation_martina}
M.~S. Karl, ``Unraveling the origin of high-energy neutrino sources: follow-up
  searches of icecube alert events,'' 2022.
\newblock \url{https://mediatum.ub.tum.de/?id=1654955}.

\bibitem{Albert_2022}
{\bfseries ANTARES, IceCube, Pierre Auger, Telescope Array} Collaboration,
  A.~Albert, S.~Alves, M.~Andr{\'{e} }, M.~Anghinolfi, M.~A., {\em et~al.}
  \href{http://dx.doi.org/10.3847/1538-4357/ac6def}{{\em The Astrophysical
  Journal} {\bfseries 934} no.~2, (Aug, 2022) 164}.

\end{thebibliography}\endgroup
